# DEVELOPMENT OF HOLLOW ELECTRON BEAMS FOR PROTON AND ION COLLIMATION*

G. Stancari,† A. I. Drozhdin, G. Kuznetsov, V. Shiltsev,
D. A. Still, A. Valishev, L. G. Vorobiev (FNAL, Batavia, IL 60510, USA),
R. Assmann (CERN, Geneva, Switzerland), A. Kabantsev (UCSD, La Jolla, CA 92093, USA)

*Abstract*

Magnetically confined hollow electron beams for controlled halo removal in high-energy colliders such as the Tevatron or the LHC may extend traditional collimation systems beyond the intensity limits imposed by tolerable material damage. They may also improve collimation performance by suppressing loss spikes due to beam jitter and by increasing capture efficiency. A hollow electron gun was designed and built. Its performance and stability were measured at the Fermilab test stand. The gun will be installed in one of the existing Tevatron electron lenses for preliminary tests of the hollow-beam collimator concept, addressing critical issues such as alignment and instabilities of the overlapping proton and electron beams.

## HOLLOW-BEAM COLLIMATOR

A collimation system must protect equipment from intentional and abnormal beam aborts by intercepting particle losses. Its functions also include controlling and reducing the beam halo, which is continually replenished during normal operations by various processes such as beam-beam collisions, intrabeam scattering, beam-gas scattering, rf noise, ground motion, and betatron resonances. Uncontrolled losses of even a small fraction of the circulating beam can damage sensitive components, quench superconducting magnets, or produce intolerable experimental backgrounds. Conventional collimation schemes are based on collimators and absorbers, possibly incorporating several stages. In the Tevatron, the primary collimators are 5-mm tungsten plates positioned about 5 standard deviations ($\sigma$) away from the center of the beam core. The absorbers (or secondary collimators) are 1.5-m steel jaws at $6\sigma$. In the LHC, the primaries are 0.6-m carbon jaws at $6\sigma$, whereas the 1-m carbon/copper secondaries are positioned at $7\sigma$ [1]. At present, there is no viable solution for scraping the beam tails in the LHC at full intensity (350 MJ per beam), as no material can be brought closer than about $5\sigma$. This project is devoted to the study of hollow electron beams as a possible candidate for collimation of high-intensity proton or ion beams.

* Fermilab is operated by Fermi Research Alliance, LLC under Contract No. DE-AC02-07CH11359 with the United States Department of Energy. This work was partially supported by the United States Department of Energy through the US LHC Accelerator Research Program (LARP).

† On leave from Istituto Nazionale di Fisica Nucleare, Sezione di Ferrara, Italy. E-mail: stancari@fnal.gov.

The hollow electron beam collimator (HEBC) is a magnetically confined, optionally pulsed electron beam with a hollow current-density profile overlapping with the proton or ion beam of interest [2]. The core passes through the center of the electron distribution and is unperturbed. The halo experiences nonlinear transverse kicks and it is driven towards the collimators. The electron gun is immersed in a conventional solenoid and provides a few amperes of current at 10 keV. The overlap region is contained within the cryostat of a superconducting solenoid providing an axial field of up to 6 T. The electron beam is then driven towards a water-cooled collector inside a separate conventional solenoid. The electron beam can be placed close to an intense beam core without any material damage.

In cylindrical symmetry, the kicks experienced by protons traversing an electron beam at a radius $r$ encompassing a current $I_r$ in an interaction region of length $L$ are given by the following expression:

$$\theta_r = \frac{2 I_r L (1 \pm \beta_e \beta_p)}{r \beta_e \beta_p c^2 (B\rho)_p} \left( \frac{1}{4\pi\epsilon_0} \right),$$

where $\beta_e c$ is the electron velocity, $\beta_p c$ the proton velocity, and $(B\rho)_p$ is the magnetic rigidity of the proton beam. The $+$ sign applies when the magnetic force is directed like the electrostatic attraction ($\mathbf{v}_e \cdot \mathbf{v}_p < 0$). For example, in a configuration similar to a Tevatron electron lens [3] ($I_r = 2.5$ A, $L = 2$ m, $\beta_e = 0.19$, $r = 3.5$ mm), the corresponding kicks are 2.4 $\mu$rad for 150-GeV protons and 0.36 $\mu$rad at 980 GeV. The r.m.s. kicks due to multiple Coulomb scattering in a Tevatron collimator are 110 $\mu$rad at 150 GeV and 17 $\mu$rad at 980 GeV. At 7 TeV, the LHC collimators impart an r.m.s. kick of 4.5 $\mu$rad. Large electron currents ($\sim$50 A) would be necessary for the HEBC to provide the same kicks as a conventional collimator and clean halo particles in a few revolutions. One important difference between the HEBC and conventional schemes is that the hollow-beam kicks are not random in space or time. The electric field is determined by the electrons' current distribution, and the electron beam can be continuous or pulsed with rise times below 100 ns. Resonant excitation tuned to a strong lattice resonance is possible. This technique is very effective, as demonstrated by calculations and by abort-gap clearing with electron lenses tuned to the 3rd and 7th order resonances in the Tevatron [4].

Analytical expressions for the current distribution were used to estimate the effectiveness of the HEBC on a proton beam. They were included in tracking codes such as STRUCT, LIFETRAC, and SixTrack [6] to follow core and halo particles as they propagate in the machine lattice [2, 7]. These codes are complementary in their treatment of apertures, field nonlinearities, and beam-beam interactions. From the measurements of the actual density profiles (described below), a 2-dimensional map of the fields was extracted using a Poisson solver. A full 3D description of the electron beam through the system is being set up using the WARP code [5]. The purpose is to take into account such effects as longitudinal profile evolution or bends in electron transport. If the latter turn out to be irrelevant, an electron-lens configuration can be adopted instead of the fully symmetrical one, greatly simplifying cathode design.

From the above considerations and calculations, the HEBC emerges as a 'soft scraper' to complement and improve a conventional collimation system. The electron beam has no hard edges, losses are gradual, and loss spikes due to beam jitter are reduced. The impact parameter on the primary collimators is increased, and they may be retracted to reduce impedance, if protection of the apparatus is not compromised. In the case of ion-beam cleaning, there is no nuclear breakup. Collimator positions are controlled by magnetic correctors, and not by mechanical means. In addition, the proposed technique is grounded in the estabilished fields of electron cooling and electron lenses.

The stability of the proton-electron system and the impedance of the HEBC need to be investigated in detail. It is known that stability is not an issue for Gaussian or flat beams in a Tevatron electron lens if the confining field is of the order of 2 T or higher. Calculations exist of the treshold for the onset of transverse-mode coupling instabilities [8]. In the case of the HEBC, the situation should be more favorable, as in the ideal case the proton beam experiences no field, but this needs to be verified experimentally.

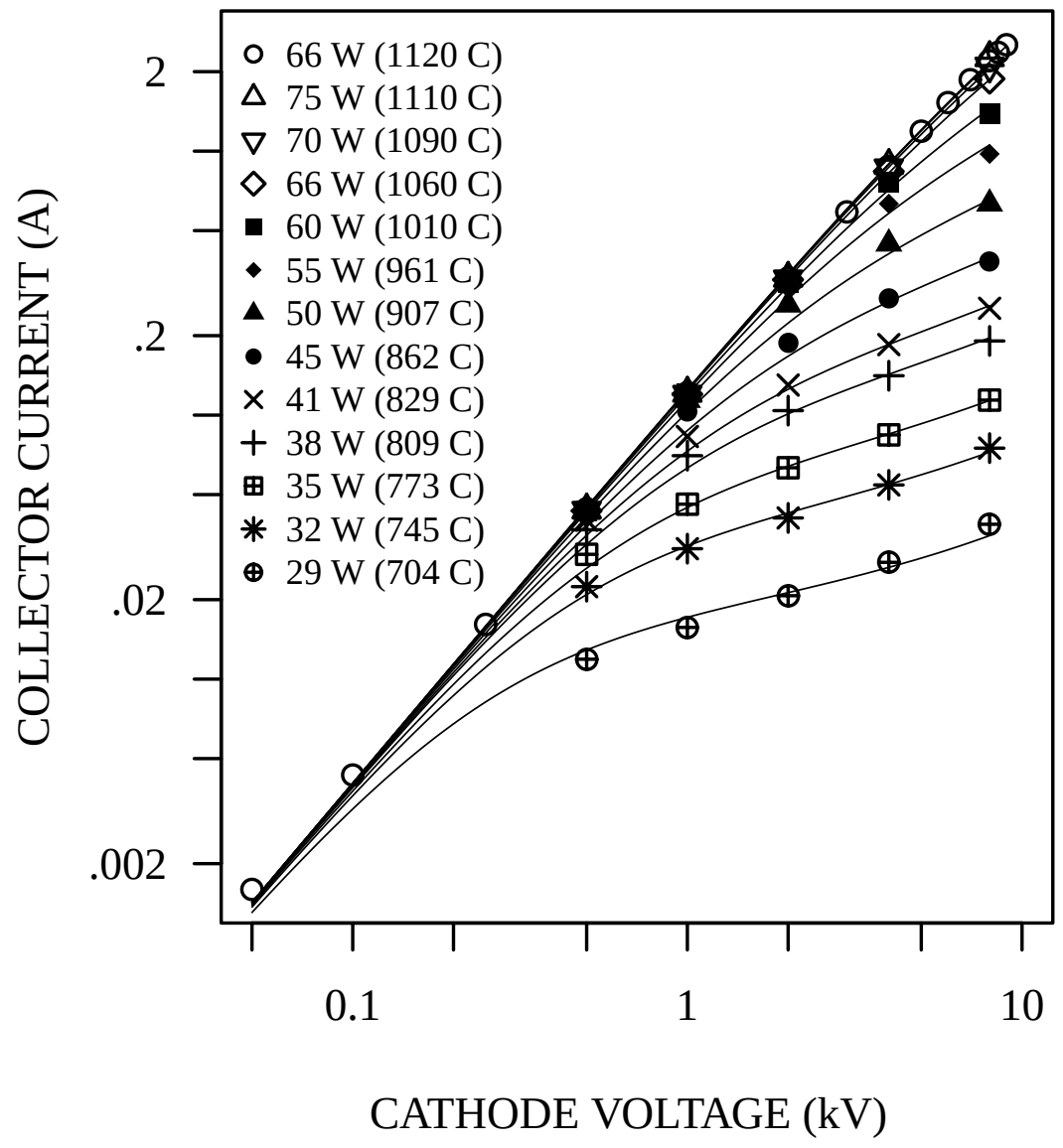

Figure 1: Performance of the 0.6-in hollow gun vs. cathode voltage and heater power.

## GUN DESIGN AND PERFORMANCE

A high-perveance gun was built to test hollow-beam collimation in the Tevatron. The present design is based upon the cathodes already employed in the Tevatron electron lenses, which are now commercially available from HeatWave Labs, Inc. (Watsonville, CA, U.S.A.). A convex tungsten dispenser cathode with BaO:CaO:$Al_2O_3$ impregnant is used to obtain high perveance [9]. The cathode has an outer diameter of 15.24 mm (0.6 in) and a radius of curvature of 10 mm. In this design, a 9-mm-diameter hole is bored along the cathode's axis. The expected profile in the space-charge-limited regime was calculated using the UltraSAM code [10]. The calculated current density distribution vanishes between $0 < r < 4.5$ mm, then rises sharply and gradually goes back to zero at the outer edges. The gun was designed at Fermilab, manufactured by Hi-Tech Manufacturing, LLC (Schiller Park, IL, U.S.A.), and installed in the Fermilab electron-lens test bench for characterization.

The test bench includes a cathode filament heater and a high-voltage modulator (10 kV maximum, 2.5 mA average current, 6 $\mu$s typical pulse width). The 3-m-long straight beamline is equipped with pickup electrodes. Three 0.4-T conventional solenoids instrumented with magnetic correctors provide independently controlled axial fields in the gun, central, and collector regions. The central solenoid was designed for electron cooling and has high field quality and low field-line ripple. The water-cooled collector has a 0.2-mm-diameter pinhole for current-density profile measurements. Typical vacuum is $2 \times 10^{-8}$ mbar.

The performance of the gun as a function of cathode voltage and heater power is shown in Figure 1. Points represent the measured peak collector current. Data is taken a few minutes after changing the heater setting, when the resistance of the filament has reached equilibrium. Thermalization of the whole gun takes longer, as shown by the two data sets at 66 W, taken a few hours apart. Data in the space-charge-limited regime yield a perveance of 4 $\mu$perv.

Current-density profiles are measured by recording the current through the pinhole while sweeping the electron beam with the magnetic correctors in small steps. An example of profile measurement is shown in Figure 2a. It was taken under the following conditions: heater power 66 W, cathode voltage $-0.5$ kV, pulse width 6 $\mu$s, repetition period 0.6 ms, peak current 44 mA, axial field 0.3 T in all 3 solenoids. At low currents, measured profiles agree with calculated space-charge-limited emission (Figure 2b).

Hollow beams are subject to breakup under the action of space charge in an axial magnetic field, due to the diocotron or slipping-stream instability [11]. For a fixed propagation length, profile evolution increases with current and decreases with magnetic field and voltage. As an example, Figure 3 shows how the current-density profile at the collector changes with increasing current for an axial field of 0.3 T. Emission features are extremely reproducible, even after cooling and reheating the cathode over the course of several months. Breakup is probably triggered by small gun misalignments or field asymmetries and not by cathode surface defects, as verified by observing the emission uniformity in the temperature-limited regime. Vortices are undesirable, as they result in a nonuniform electric field on axis of about 10% of the peak field in the worst case (see Figure 2c for a sample calculation). One may mitigate the effect with azimuthally segmented extraction electrodes. Although magnetic fields are constrained by the desired beam size in the overlap region, high fields may be used to freeze profile evolution, or it may be possible to take advantage of the $\mathbf{E} \times \mathbf{B}$ twist itself to smooth the distribution (Figure 3).

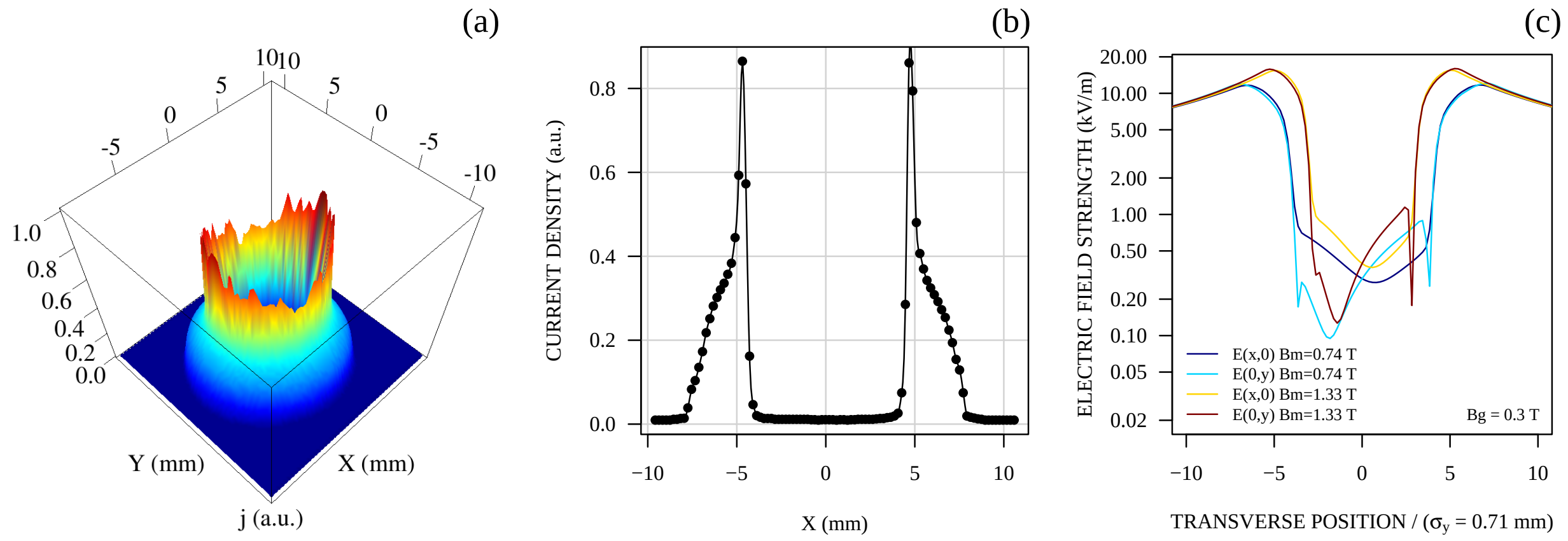

Figure 2: Measured current-density profile at 0.5 kV (a and b), and 2D WARP calculation of electric fields in TEL2 for two different values of the main solenoid field vs. transverse position in units of vertical proton beam r.m.s. size (c).

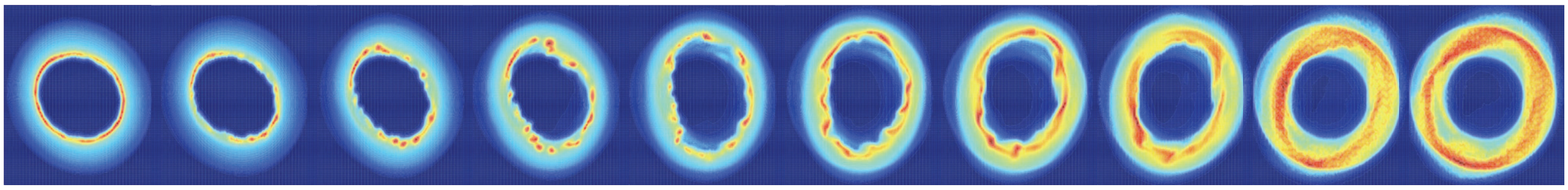

Figure 3: Measured profile evolution as voltage and current are increased from (0.5 kV, 44 mA) to (9 kV, 2.5 A).

We plan to install the 0.6-in hollow gun in one of the existing Tevatron electron lenses (TEL2). Core lifetimes, losses and loss spikes at collimators and detectors will be monitored as a function of HEBC parameters (position, angle, intensity, magnetic field, timing) for individual proton or antiproton bunches. Although kicks will be smaller than at injection, measurements will be done at flattop (980 GeV), where orbits and emittances are stable and the collimation system is well understood. Alignment procedures have been tested with the Gaussian gun currently installed in TEL2. They are based on improved beam-position monitors and on the observation of loss patterns as the electrons are scanned across the circulating beam. Installation is scheduled for the upcoming summer shutdown of the accelerator complex, and experiments can reasonably start in the fall.

We wish to thank G. Annala (FNAL), W. Fischer (BNL), N. Mokhov (FNAL), and J. C. Smith (SLAC) for discussions and insights.